\documentclass{svproc}
\usepackage{graphicx}
\usepackage{array}
\usepackage{url}

\begin{document}
\mainmatter              
\title{Brain Tumor Classification on MRI in Light of Molecular Markers}
\titlerunning{Brain Tumor classification}  
%
\author{Jun Liu\inst{1,2}\and Geng Yuan\inst{2}\and Weihao Zeng\inst{1}\and Hao Tang\inst{1}\and Wenbin Zhang\inst{3}\and Xue Lin\inst{2}\and XiaoLin Xu\inst{2}\and Dong Huang\inst{1}\and Yanzhi Wang\inst{2}}
\authorrunning{Jun Liu et al.} 
%
\tocauthor{Jun Liu, Geng Yuan, Weihao Zeng, Hao Tang, Wenbin Zhang,
Xue Lin, XiaoLin Xu, Dong Huang and Yanzhi Wang}
\institute{Robotics Institute, School of Computer Science, Carnegie Mellon University, Pittsburgh, PA, USA
\and
Department of Electrical \& Computer Engineering, College of Engineering, Northeastern University, Boston, MA, USA
\and
University of Maryland, Baltimore County, MD, USA}

\maketitle              

\begin{abstract}
In research findings, co-deletion of the 1p/19q gene is associated with clinical outcomes in low-grade gliomas. The ability to predict 1p19q status is critical for treatment planning and patient follow-up. This study aims to utilize a specially MRI-based convolutional neural network for brain cancer detection. Although public networks such as RestNet and AlexNet can effectively diagnose brain cancers using transfer learning, the model includes quite a few weights that have nothing to do with medical images. As a result, the diagnostic results are unreliable by the transfer learning model. To deal with the problem of trustworthiness, we create the model from the ground up, rather than depending on a pre-trained model. To enable flexibility, we combined convolution stacking with a dropout and full connect operation, it improved performance by reducing overfitting. During model training, we also supplement the given dataset and inject Gaussian noise. We use three--fold cross-validation to train the best selection model. Comparing InceptionV3, VGG16, and MobileNetV2 fine-tuned with pre-trained models, our model produces better results. On a validation set of 125 codeletion vs. 31 not codeletion images, the proposed network achieves 96.37\% percent F1-score, 97.46\% percent precision, and 96.34\% percent recall when classifying 1p/19q codeletion and not codeletion images.
\end{abstract}

\section{Introduction}

Tumors of the brain are malignant cell growths or aggregates in or around the brain.Depending on where they exist in the brain, low-grade gliomas~\cite{akkus2017predicting,bhattacharya2021determining} can manifest themselves in a variety of ways. The patient may have weakness or numbness in the right leg if the brain cancer grows in the area of the brain that controls it~\cite{amin2022brain}. 

The most efficient method to detect brain tumors is MRI~\cite{liu2021explainable,liu2022efficient,liu2023interpretable}. Scanning generates a vast amount of magnetic resonance images, which is examined by a radiologist. Biomarker detection helps give patients the best appropriate treatment for their particular condition. This study is notable for the unique and promising findings of merging deep learning with radiogenomics. Deep learning was better at detecting 1p/19q co-deletion on T2 images than it does on post-T1 contrast images.

Akkus et al. ~\cite{akkus2017predicting} were the first to use a deep learning approach to predict 1p19q from low-grade glioma MRI images in 2017. Chelghoum et al., 2020~\cite{chelghoum2020transfer} employs transfer learning to classify 1p19q using popular pre-trained models such as AlexNet, VGG19, GoogleNet, and others.\cite{lombardi2020clinical,6,7} They point out that transfer learning can still provide correct results even with limited datasets. To get the best accuracy, Abiwinanda et al. \cite{8} create the network by combining various CNN operations. Maithra Raghu et al. \cite{9} discovered that transfer learning had little impact on imaging tasks in medicine, with models trained from scratch performing nearly as well as ImageNet-transferred models.

Our contributions are listed below.
\begin{itemize}
\item Using a convolution stack, we develop a CNN specifically for detecting brain cancer in MRI images.
\item To avoid overfitting and improve performance, we utilize a tunable composition of dropout and Gaussian noise during training.
\item Comprehensive evaluation of discriminant results using the confusion matrix, F1-score, precision, and recall methods to evaluate unbalanced data in order to avoid false positives.
\end{itemize}

\section{Materials and Methods}

To train our proposed network, we use the public Kaggle dataset\cite{10}. Meanwhile, in this data set, we compare the results of VGG16, InceptionResNetV2 and MobileNetV2, all fine-tuned~\cite{liu2025rora} pre-trained models applying transfer learning.

\subsection{Experimental Data}

For evaluation and research, Kaggle Public Datasets offers brain MRI datasets. There are 253 brain Magnetic Resonance images in the collection, divided into two folders: Yes and No. The folder Yes contains 155 scans of tumors in the brain, while the folder No has 98 non-tumor brain MRI scans.
\begin{figure}
\centering
\includegraphics[height=0.2\textheight, width=0.60\textwidth]{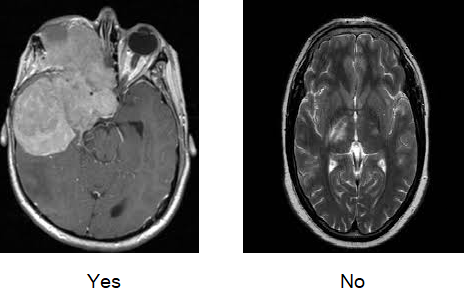}
\caption{Brain Magnetic Resonance Image}\label{fig1}
\end{figure}

A brain with a tumor is on the left, while a healthy brain is on the right in Fig. \ref{fig1}.

\newpage
\subsection{Network Composition}

\begin{figure}
\centering
\includegraphics[width=\textwidth]{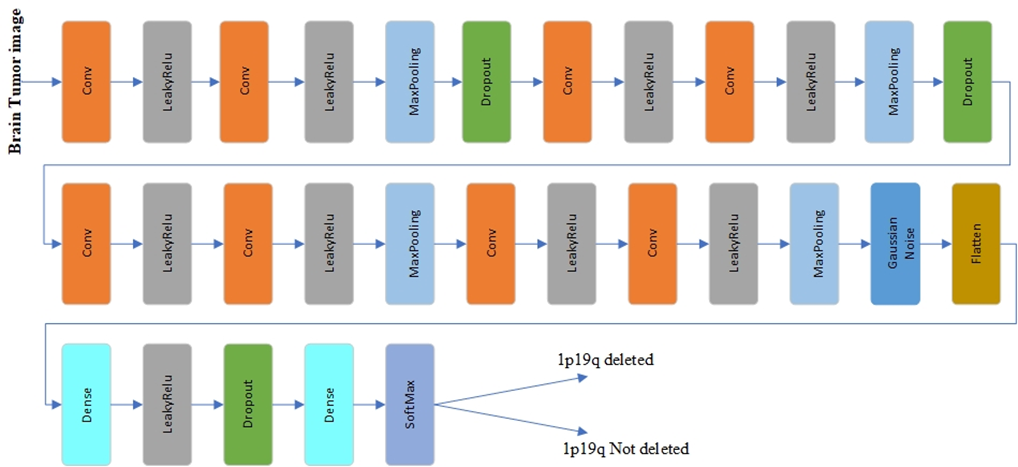}
\caption{Network Architectures}\label{fig2}
\end{figure}

The model seen in Fig. \ref{fig2} has 14 layers. Kernels with convolutions (3 x 3) had positive results, because the small convolutions catch some of the finer details of the edges. It starts with 16 kernels in first two CNN layers and gradually progresses to 32, 64, and lastly 128 kernels per layer.
  
Convolution layers, pooling layers, LeakyRelu layer, Softmax layer, dropout layers,  Dense layers\cite{11}, Flatten layer,  make up the network represented in Fig.\ref{fig2} .

\begin{table}[tb!]
\caption{~}\label{tab1}
\begin{center}
\begin{tabular}{m{2.5cm}m{1.5cm}m{1.5cm}m{1.5cm}m{2.5cm}m{2cm}<{\centering}m{2cm}<{\centering}}
\hline\rule{-3pt}{12pt}
Layer &  Kernels & Kernel Size & Stride & Feature Map Size & Activation\\[2pt]
\hline\rule{-3pt}{12pt}
InputLayer & - & - & - & 256 x 256 & -  \\
Convolution & 16 & 3 x 3 & 1 & 254 x 254 x 16  & LeakyReLU\\
Convolution &  16 & 3 x 3 &1 & 252 x 252 x 16 & LeakyReLU\\
MaxPooling & - & 2 x 2 & 1 & 126 x 126 x 16 & -\\
Dropout &  - & - & -& 126 x 126 x 16  & -\\
Convolution &  32  & 3 x 3 & 1& 124 x 124 x 32 & LeakyReLU\\
Convolution & 32  & 3 x 3 & 1 & 122 x 122 x 32 & LeakyReLU\\
MaxPooling & - & 2 x 2 & 1 & 61 x 61 x 32 & -\\
Dropout &  - &  - &  - & 61 x 61 x 32 & -\\
Convolution &  64 & 3 x 3 & 1 & 59 x 59 x 64 & LeakyReLU\\
Convolution &  64 & 3 x 3 & 1 & 57 x 57 x 64 & LeakyReLU\\
MaxPooling & - & 2 x 2 & 1 & 28 x 28 x 64 & -\\
Convolution &  128 & 3 x 3 & 1 & 26 x 26 x 128 & LeakyReLU\\
Convolution &   128 & 3 x 3 & 1 & 24 x 24 x 128 & LeakyReLU\\
MaxPooling & - & 5 x 5 & 1 & 4 x 4 x 128 & -\\
GaussianNoise & - & - & - & 4 x 4 x 128 & -\\
Flatten &  - & -& - &2048 & -\\
FullConnect &  - &- & - & 1024 & -\\
Dropout   & -   & - & - & 1024 & -\\
FullConnect   & - & - & - & 2 & -\\
Softmax  & - & - & - & 2 & -\\[2pt]
\hline
\end{tabular}
\end{center}
\end{table}

The composition of the network, including the Kernel Size,stride ,Feature Map Size etc, are listed in Table \ref{tab1}.
\begin{itemize}
\item All brain tumor images supplied into the network are resized to 256 by 256 pixels.
\item For all convolutional layers, the network employs kernels of size 3 x 3, with 16, 32, 64, and 128 kernels in each layer in turn. 
\item LeakyReLU was used as the activation function since negative numbers are preserved together with concerns about saturation are eliminated when using tanh. 
\item This model starts with 2 x 2 maxpooling and subsequently progresses to 7 x 7 maxpooling. 
\item The Full connect layer is utilized twice to decrease the quantity  of neurons. 
\item On purpose to reduce overfitting, Gaussian Noise was added in the training process and it can be thought of as a method of random data enrichment.
\item Full Connected's purpose is to reduce multi-dimensional inputs to a single dimension.
\item In the course of network training, some neural network units are dropped from the network with a certain probability.
\end{itemize}

\subsection{Hyperparameters setting}

Some hyperparameter settings were investigated during this research.

\noindent\textbf{Learning rate}

Since initial weight values are relatively random, starting with a higher learning rate usually works alright. As the training phase advances, the findings often get nearer either to global or local minima. 

\noindent\textbf{EarlyStopping}

The model is prevented from overfitting the training data by using early stop approaches. The model will be stopped from training if there is no change of at least 0.001 in 8 epochs.

\noindent\textbf{Batch Size}

The batch size is limited by the amount of RAM available. Moreover, although a larger batch size helps for weights update less frequently, faster training,  which may result in less effective results.

\noindent\textbf{Count of Epochs}

The number of epochs indicates how many times the entire training data is iterated over when training the model.

\subsection{Experiments}

We develop with Keras within the TensorFlow framework in this experiment.

\noindent\textbf{Data preprocessing}

The data preprocessing steps are the removal of the third class, normalization of the data, and reshuffling of the training data. Since there is no enough image to train the model with this small dataset, so data augmentation aids in addressing the data imbalance issue.

\noindent\textbf{Proposed training procedure}

\begin{figure}[tb!]
\centering
\includegraphics[width=\textwidth]{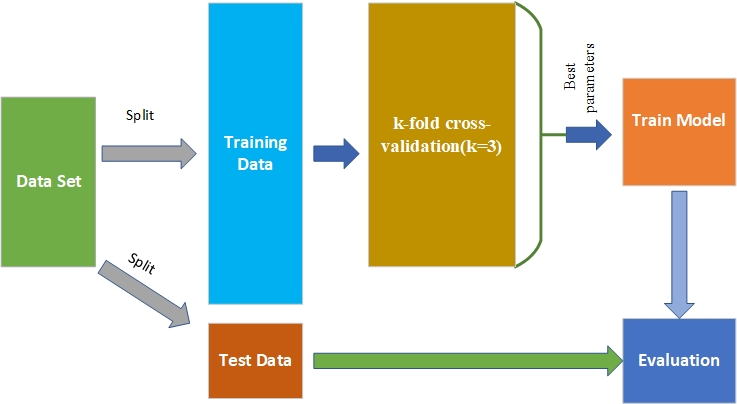}
\caption{Proposed training procedure}\label{fig3}
\end{figure}

In order to improve performance, we develop, evaluate, and train our model, which is represented in Fig. 
~\ref{fig3} using cross-validation in model training.

\begin{itemize}
\item The data set is separated into train and test sets at stochastic, with the model was build using the train set, while the test set was utilized to assess its accuracy. 
\item The program fine-tunes the model by cross-validation with k-fold \cite{12} to obtain the finest quality model.
\item Assess the model's anticipated accuracy on the test set.
\end{itemize}

\noindent\textbf{Evaluation method}

To evaluate our designed model, we use the confusion matrix, accuracy, precision, recall, and F1.

\noindent\textbf{Confusion Matrix}

A method for analyzing the performance of classification algorithms is called the confusion matrix. If the dataset have an unbalanced amount of observations in each class or if the dataset has more than two classes, it would be incorrect to use classification accuracy alone as a measurement.

\begin{table}
\caption{Confusion  Matrix}\label{tab2}
\begin{center}
\begin{tabular}{m{3cm}m{3cm}<{\centering}m{3cm}<{\centering}}
\hline\rule{-3pt}{12pt}
 & 1p/19q deleted & 1p/19q not deleted\\[2pt]
\hline\rule{0pt}{12pt}
1p19q deleted & True Positive & False Positive\\
1p19q not deleted &  False Negative & True Negative\\[2pt]
\hline
\end{tabular}
\end{center}
\end{table}

Table \ref{tab2} explicitly displays the proportion of accurate and wrong identifications for each category.

\begin{figure}[!h]
\centering
\includegraphics[width=0.7\textwidth]{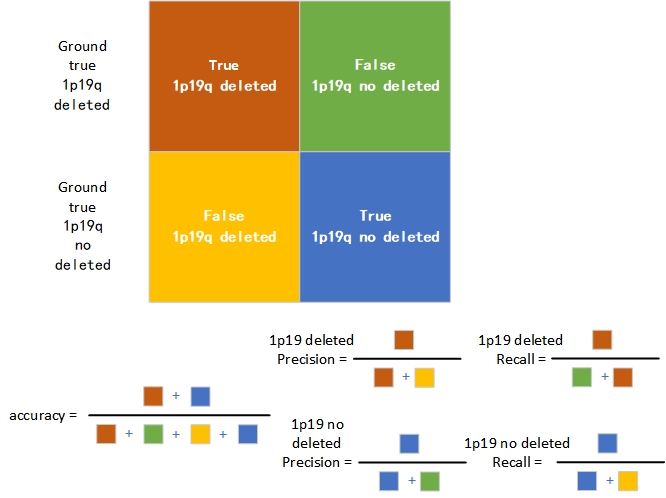}
\caption{Evaluation matrix}\label{fig4}
\end{figure}
From Fig. \ref{fig4} most metrics that can be derived from the confusion matrix.

\noindent\textbf{Precision}

Precision refers to the percentage of 1p/19q deleted sufferers correctly predicted by the model terms of number of sufferers with 1p/19q deleted.
\begin{center}
Precision = True 1p/19q deleted / (True 1p/19q deleted  + False 1p/19q deleted)
\end{center}

\noindent\textbf{Recall}

The percentage of 1p/19q deleted sufferers  detected by the model to all 1p/19q deleted sufferers is used to calculate specificity.
\begin{center}
Recall = True 1p/19q deleted / (True 1p/19q deleted  + False 1p/19q non-deleted)
\end{center}

\noindent\textbf{F1 score}

The F1 score was established to work successfully with unbalanced data because of a disparity in the percentage of brain and non-brain malignancies. in this dataset. Its formula is as follows:

\begin{center}
F1score = 2 * (recall * precision) / (recall + precision)
\end{center}

\section{Results}

\begin{table}[!h]
\caption{Result}\label{tab3}
\vspace{-3mm}
\begin{center}
\setlength\tabcolsep{5pt}
\begin{tabular}{lcccc}
\hline\rule{-3pt}{12pt}
type & Precision & Recall & F1-score & Images\\[2pt]
\hline\rule{-3pt}{12pt}
1p/19q deleted & 0.9881 & 0.9635 & 0.9742 & 125\\
1p/19q not deleted & 1 & 0.9644 & 0.9375 & 31\\
avg / total & 0.9746 & 0.9634 & 0.9637 & 156\\[2pt]
\hline
\end{tabular}
\end{center}
\vspace{-5mm}
\end{table}

The suggested network achieves a f1-score of 0.9637 in Table \ref{tab3} using 164 images, including 125 1p19q deleted and 39 1p19q undeleted.

\begin{figure}[tb!]
\centering
\includegraphics[width=0.45\textwidth]{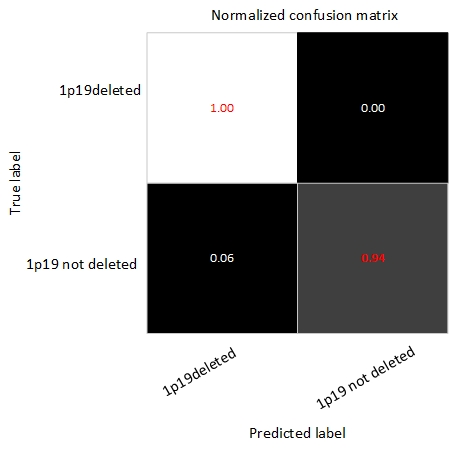}
\caption{Confusion matrix}\label{fig5}
\vspace{-2mm}
\end{figure}

On the test set, Fig. \ref{fig5} illustrates the 1p19q values of the confusion matrix. We can observe that all 125 1p19q deleted images have been correctly identified.

To compare with our model, we used transfer learning based models such InceptionResNetV2, MobileNetV2, and VGG16.

\begin{table}]
\caption{Baseline Comparisons}\label{tab5}
\begin{center}
\setlength\tabcolsep{5pt}
\begin{tabular}{lccc}
\hline\rule{-3pt}{12pt}
Model & Precision & Recall & F1 score\\[2pt]
\hline\rule{-3pt}{12pt}
We proposed & 0.9746 & 0.9634 & 0.9637\\
InceptionV3\cite{13} & 0.9230 & 1 & 0.9600\\
MobileNetV2\cite{14} & 1.0 & 0.8709 & 0.9200\\
InceptionResNetV2\cite{14} & 0.9062 & 0.9354 & 0.9153\\
VGG16\cite{ruslankl} & 0.8960 & 0.9286 & 0.9123\\[2pt]
\hline
\end{tabular}
\end{center}
\vspace{-5mm}
\end{table}

From Table \ref{tab5}, we observe that our model is well-balanced in terms of Precision, Recall, and F1 score, with these values being very close to each other. The F1 score, as a comprehensive metric that combines Precision and Recall, provides a clearer picture of the model's performance in recognition. In terms of performance, our model surpasses all other models in terms of the F1 score.

\section{Discussion}
We present a viable CNN-based technique for predicting deletions of the 1p/19q chromosomal arm. A significant challenge in applying deep learning algorithms to medical imaging is the lack of sufficient datasets. For medical diagnosis, current transfer learning methods rely on various publicly available models trained on large ImageNet datasets. However, these models often generate a substantial number of medical images, which can compromise the accuracy of clinical diagnosis~\cite{chinta2025ai,kuraparthi2021brain}.

Transfer learning is not used in our brain tumor detection model, and the parameters it creates are all based on medical imaging datasets, ensuring complete accuracy in the brain tumor detection.

A small dataset~\cite{liu2022efficient} can result in substantial training errors. In order to solve this problem, Gaussian noise is added during the training phase to improve the model's normalization ability and fault tolerance. Adding noise prevents the network from memorizing the training samples, as they are constantly changing. As a result, the network weights are reduced, making the network more robust and leading to a lower generalization error. Since fresh samples are drawn close to existing samples in the input space, the shape of the input space is smoothed out. This makes it easier for the network to learn the mapping function, resulting in faster and more effective learning.
In the next step, we will explore using diffusion~\cite{meng2024instructgie} models to generate small samples, combine them with LLM for 3D reconstruction~\cite{lei2023mac,dong2024physical,dong2024df}, and apply compression~\cite{liu2025toward,li2025mutual,limutual,tan2025harmony} methods for deployment on embedded devices~\cite{niu2025mobile,yuan2021work,yuan2022mobile,liu2024tsla,ji2025computation,liu2023scalable}.
\section{Conclusion}

In this paper, we use convolution stacking in conjunction with Gaussian noise and random drop in training to create a model for identifying a small dataset of brain tumors that outperforms the transfer learning method while having the advantages of small model size and high reliability. Cross-validation in training is an excellent way to overcome the problem of data imbalance.

\section{ACKNOWLEDGMENT}
Thanks to NVIDIA for their technical assistance.

%
%

\end{document}